\documentclass[usenatbib]{mn2e}
\usepackage{amsmath,graphics,amssymb}
\usepackage{times}

\allowdisplaybreaks

\newcommand{\de}{\mathrm{d}}
\newcommand{\Teff}{\mbox{$T_\mathrm{eff}$}}
\newcommand{\zav}[1]{\left(#1\right)}
\newcommand{\hzav}[1]{\left[#1\right]}

\DeclareMathAlphabet{\mathsc}{OT1}{cmr}{m}{sc}
\def\testbx{bx}%
\DeclareRobustCommand{\ion}[2]{%
\relax\ifmmode
\ifx\testbx\f@series
{\mathbf{#1\,\mathsc{#2}}}\else
{\mathrm{#1\,\mathsc{#2}}}\fi
\else{#1\,{\scshape{#2}}}%
\fi}

\title[Influence of XUV radiation on \ion{P}{v} ionization fraction in hot
      star winds]
      {Influence of XUV radiation on \ion{P}{v} ionization fraction in hot
      star winds}

\author[J.  Krti\v{c}ka and J. Kub\'at]{Ji\v r\'\i\  Krti\v{c}ka$^{1}$\thanks{E-mail: krticka@physics.muni.cz (JKr); kubat@sunstel.asu.cas.cz (JKu)
	} and Ji\v r\'\i\  Kub\'at$^{2}$%
\footnotemark[1]\\
	$^{1}$\'Ustav teoretick\'e fyziky a astrofyziky,
        Masarykova univerzita,
	CZ-611 37 Brno, Czech Republic\\
	$^{2}$Astronomick\'y \'ustav
	AV \v{C}R, Fri\v{c}ova 298,
	CZ-251 65 Ond\v{r}ejov, Czech Republic}

\begin{document}

\date{Received}

\maketitle

\begin{abstract}

Different diagnostics of hot star wind mass-loss rates provide results that are
difficult to reconcile with each other. The widely accepted presence of clumping
in hot star winds implies a significant reduction of observational mass-loss
rate estimates from diagnostics that depend on the square of the density.
Moreover, the ultraviolet \ion{P}{v} resonance lines indicate a possible need
for even stronger reduction of hot star mass-loss rates, provided that
\ion{P}{v} is a dominant ionization stage of phosphorus at least in some hot
stars. The latter assumption is challenged by a possible presence of the XUV
radiation.

Here we study the influence of the XUV radiation on the \ion{P}{v} ionization
fraction in the hot star winds. By a detailed solution of the hydrodynamical,
radiative transfer, and statistical equilibrium equations we confirm that
sufficiently strong XUV radiation source may decrease the \ion{P}{v} ionization
fraction, possibly depreciating the \ion{P}{v} lines as a reliable mass-loss
rate indicator. On the other hand, the XUV radiation influences also the
ionization
fraction of heavier ions that drive the wind, leading to a decrease of the wind
terminal velocity. Consequently, we conclude that the XUV radiation alone can
not bring theory and observations in accord.

We fit our predicted wind mass-loss rates by a suitable formula and compare the
results with the observational mass-loss rate diagnostics. We show that for
supergiants and giants the theoretical predictions do not contradict the
mass-loss rate estimates based on X-ray line profiles or density squared
diagnostics. On the other hand, for main-sequence stars the predicted mass-loss
rates are still significantly higher than that inferred from \ion{P}{v} or X-ray
lines. This indicates that the "weak wind problem" recently detected in
low-luminosity main-sequence stars may occur to some extent
also for the stars with higher luminosity. 

\end{abstract}



\section{Introduction}

Mass-loss plays an important role in the massive star evolution. Most of the
mass of massive stars is lost during their evolution from zero-age main sequence
to the final remnant. Different processes contribute to the mass loss in
individual evolutionary phases. These processes include line-driven winds during
hot evolutionary stages \citep{pulvina}, decretion disks in fast-rotating stars
\citep{los91,sapporo}, LBV-type of explosions in hot supergiants
\citep{vodovar}, dust-driven winds in cool supergiants \citep{wojta}, and a
final supernova explosion \citep{medaci}. Consequently, estimates of amount of
mass lost per unit of time (mass-loss rate) as a functions of stellar parameters
belong to one of the most important ingredients of evolutionary models.

Unfortunately, the uncertainties in modern mass-loss rate determinations
significantly affect the evolutionary models of massive stars. In the case of
the line-driven wind of hot stars, these uncertainties seem to be mostly
connected with the occurrence of small-scale inhomogeneities \citep{potclump}.
The inhomogeneities are typically divided into three differents groups
(microclumping, porosity, and vorosity) according to their influence on the
spectral features, although they may be caused by the same structure observed in
different wavelengths. Microclumping (also frequently referred to as clumping),
that accounts for the enhanced density in the optically thin inhomogeneities,
can be most easily incorporated in the wind models
\citep[e.g.,][]{hagr,pulchuch,sanya}. Microclumping affects the ionization
equilibrium via enhanced recombination, consequently it influences the radiative
transfer only indirectly \citep{abbob}. On the other hand, porosity (also
referred to as macroclumping) accounts for the nonnegligible optical depth of
inhomogeneities (which may become optically thick), and directly infuences the
radiative transfer \citep{lidaarchiv,sund,clres1}. Vorosity affects the line
profiles, and it is connected with the different Doppler shifts of individual
inhomogeneities \citep{ocvor}. From the point of view of mass-loss rate
predictions, the inhomogeneities may affect the ionization fractions of wind
driving ions \citep{sanya,muij}, or they may lead to the decrease of the wind
mass-loss rate due to the base turbulence \citep{lucyinthewind,cmf1}.

There could be a simple observational solution of the mass-loss rate
determination problem: to find such observational characteristic that is not
affected by the wind inhomogeneities. There are two potential candidates for
such convenient observables: the X-ray
radiation \citep[either line profiles or the continuum
flux distribution,]
[]{mcf,owco,tlustocerv,coh},
and unsaturated resonance line profiles. The latter
case is fulfilled for trace elements, and especially interesting is the
\ion{P}{v} ion \citep{prvnifosfor,maso,full}. However, even these
characteristics face some problems. The opacity in the X-ray domain scales
mostly linearly with the density, consequently its effect on the X-ray
diagnostics
should be in principle possible to model in a straightforward way.
Indeed, the observations of mostly symmetric X-ray line profiles, that are not
strongly affected by absorption, indicate low wind mass-loss rates
\citep{celakoza,cohzet,gagne}. These results are, however, challenged by a
possible effect of the  porosity on the X-ray opacity \citep{lidarikala,oski}.
Note that there is not a general consensus on this problem
\citep{taulida,tauowo}.

Here we concentrate mostly on the other promising observational characteristic,
i.e., on the \ion{P}{v} resonance lines. The weakness of observed \ion{P}{v}
lines also indicates low mass-loss rates of hot star winds \citep{full}.
However, the ionization fraction of \ion{P}{v} may be modified by the effect of
clumping \citep{pulamko,sanya}. Additional changes of the \ion{P}{v} ionization
fraction may be caused by the influence of X-rays. Although our previous
calculations indicated that \ion{P}{v} ionization fraction is not strongly
affected by X-rays \citep{nlteiii}, \citet{walcp} argued that the extreme
ultraviolet radiation (hereafter XUV) may affect \ion{P}{v} ionization
fractions. Since the work of \cite{walcp} was based on rather simplified
ionization estimates and the calculations presented in \citet{nlteiii} were done
without enhanced XUV radiation, we decided to fill this gap and we apply our
NLTE wind models to study the influence of the XUV radiation (parameterized in a
convenient way) on the \ion{P}{v} ionization fraction. The XUV region is defined
here as the energy from interval 54.4\,eV (\ion{He}{ii} edge) to 124\,eV.

\section{Wind models}

For our calculations we use NLTE wind models of \citet{cmf1} with a comoving
frame (CMF) line force. Our models assume stationary, and spherically symmetric
wind flow. They enable us to selfconsistently predict wind structure just from
the stellar parameters (the effective temperature, mass, radius, and chemical
composition). The line radiative force is calculated directly by summing the
contribution from individual atomic transitions, i.e. we do not use the CAK line
force parameters.

The ionization and excitation state of the considered elements is derived from
the statistical equilibrium (NLTE) equations. Ionic models are either adopted
from the TLUSTY grid of model stellar atmospheres \citep{ostar2003,bstar2006} or
are created by us using the data from the Opacity and Iron Projects
\citep{topt,zel0}. For phosphorus we employed data described by Pauldrach et al.
(\citeyear{pahole}). Auger photoionization cross sections from individual
inner-shells were taken from  \citet[see also
\citealt{muzustahnouthura}]{veryak}, and Auger yields were taken from
\citet{kame}. The emergent surface flux is taken from H-He spherically symmetric
NLTE model stellar atmospheres of \citet[and references therein]{kub}. For our
wind calculations we assume a solar chemical composition after \citet{asp09}.

The radiative force is calculated using the solution of the spherically
symmetric CMF radiative transfer equation \citep{mikuh}. The corresponding line
data were extracted in 2002 from the VALD database (Piskunov et al.
\citeyear{vald1}, Kupka et al. \citeyear{vald2}). The radiative cooling and
heating terms are derived using the electron thermal balance method (Kub\'at et
al., \citeyear{kpp}). For the calculation of the radiative force and the
radiative cooling and heating terms we use occupation numbers derived from the
statistical equilibrium equations. The hydrodynamical equations, i.e., the
continuity equation, equation of motion with the CMF line force, and the energy
equation with radiative heating and cooling included are solved iteratively to
obtain the wind density, velocity, and temperature structure. The wind mass-loss
rate is derived from the critical condition \citep{cak} generalized for the case
of CMF line force. The derived mass-loss rate corresponds to the maximum one for
which smooth transonic solution can be obtained \citep{havran}.

\begin{table}
\caption{Stellar parameters of the model grid}
\centering
\label{ohvezpar}
\begin{tabular}{rccrcc}
\hline
\hline
& model &$\Teff$ & $R_{*}$ & $M$ & $\dot M$ \\
& & $[\text{K}]$ & $[\text{R}_{\odot}]$ &
$[\text{M}_{\odot}]$ & $[\text{M}_{\odot}\,\text{year}^{-1}$]\\
\hline main
& 300-5 & 30\,000 & 6.6 & 12.9 & $9.5\times10^{-9}$\\ sequence
& 325-5 & 32\,500 & 7.4 & 16.4 & $9.8\times10^{-9}$\\
& 350-5 & 35\,000 & 8.3 & 20.9 & $4.6\times10^{-8}$\\
& 375-5 & 37\,500 & 9.4 & 26.8 & $1.8\times10^{-7}$\\
& 400-5 & 40\,000 &10.7 & 34.6 & $6.4\times10^{-7}$\\
& 425-5 & 42\,500 &12.2 & 45.0 & $1.2\times10^{-6}$\\
\hline giants                                    
& 300-3 & 30\,000 &13.1 & 19.3 & $8.7\times10^{-8}$\\
& 325-3 & 32\,500 &13.4 & 22.8 & $1.8\times10^{-7}$\\
& 350-3 & 35\,000 &13.9 & 27.2 & $4.5\times10^{-7}$\\
& 375-3 & 37\,500 &14.4 & 32.5 & $1.0\times10^{-6}$\\
& 400-3 & 40\,000 &15.0 & 39.2 & $1.8\times10^{-6}$\\
& 425-3 & 42\,500 &15.6 & 47.4 & $2.9\times10^{-6}$\\
\hline supergiants                               
& 300-1 & 30\,000 &22.4 & 28.8 & $4.7\times10^{-7}$\\
& 325-1 & 32\,500 &21.4 & 34.0 & $7.7\times10^{-7}$\\
& 350-1 & 35\,000 &20.5 & 40.4 & $1.3\times10^{-6}$\\
& 375-1 & 37\,500 &19.8 & 48.3 & $2.3\times10^{-6}$\\
& 400-1 & 40\,000 &19.1 & 58.1 & $3.0\times10^{-6}$\\
& 425-1 & 42\,500 &18.5 & 70.3 & $3.7\times10^{-6}$\\
\hline
\end{tabular}
\end{table}

For our study we selected O star model grid with the effective temperatures in
the range $30\,000-4 2\,5 00\,\text{K}$. The parameters for the stars with given
effective temperatures were obtained using relations derived by \citet{okali}
for main-sequence stars, giants, and supergiants (see Table~\ref{ohvezpar}). 

Our new models predict slightly lower mass-loss rate than our older models
\citep{simx} due to inclusion of line overlaps via the solution of the CMF
radiative transfer equation \citep{cmf1}. The mass-loss rate predictions for all
models are also listed in the Table~\ref{ohvezpar}. We fitted these mass-loss
rate predictions as
\begin{equation}
\label{tricko}
\log\zav{\frac{\dot M}{10^{-6}\,{M}_\odot\,\text{year}^{-1}}}=(a+a_1
l)\log\zav{\frac{L}{10^6\,{L}_\odot}}+b+b_1l,
\end{equation}
where $l$ is the luminosity class (i.e., 1 for supergiants, 3 for giants and 5
for main-sequence stars), and
\begin{align}
\nonumber
a&=2.040, & a_1&=-0.018,\\
b&=0.552, & b_1&=0.068.
\end{align}
We compared formula Eq.~\eqref{tricko} with mass-loss rate predictions of
\citet{vikolamet} calculated for the mass-fraction of heavier elements
$Z=0.0134$ \citep[note that the solar mass-fraction of heavier elements assumed
by \citet{vikolamet} is different, $Z=0.0194$,][]{angre}. On average our models
predict slightly lower mass-loss rates by a factor of about $1.7$. Formula
Eq.~\eqref{tricko} shows an excellent agreement with \citet{cinskasmrt} models
A$^+$/A$^-$ of $\zeta$~Pup, while the predictions of
our formula are by a factor of
about $1.6$ lower than the mass-loss rate of models D$^+$/D$^-$.

\section{Additional X-ray/XUV radiation source}

We include an additional source of X-ray/XUV radiation into our wind models. For
this purpose we use the X-ray emissivity $\eta_\text{X}(r,\nu)$
\citep[][Eq.~11]{simx}, which was derived using the numerical simulations of
wind instability \citep{felpulpal}. This additional high energy emission starts
at the radius $2\,R_*$, and integrates the emission from the gas with different
shock temperatures as derived from hydrodynamical simulations. In our approach
the X-ray emission lines, which mostly contribute to the high energy emission in
hot stars, are not treated individually, but they are summed over selected
wavelengths. This approach is fully satisfactory because most of these lines are
optically thin in the cold wind, consequently the X-ray emission lines influence
the wind ionization equilibrium via Auger and direct photoionization only
\citep[e.g.,][]{mcf}.

The original expression for the X-ray emissivity $\eta_\text{X}(r,\nu)$ is
modified to test the influence of the XUV radiation on the wind ionization. We
introduce a nondimensional free parameter $\beta_\text{XUV}$ that scales the
original X-ray emissivity $\eta_\text{X}(r,\nu)$ in the XUV region. This region
includes frequencies $\nu_\text{XUV}^{\min}\le\nu\le\nu_\text{XUV}^{\max}$,
where $\nu_\text{XUV}^{\min}$ is the \ion{He}{ii} ionization threshold
(54.4\,eV), and we selected
$\nu_\text{XUV}^{\max}=3\times10^{16}\,\text{s}^{-1}$, which corresponds to
124\,eV. The modified X-ray emissivity included in our models is therefore given
by
\begin{equation}
\tilde\eta_\text{X}(r,\nu)=\left\{
\begin{array}{cc}
0, & \nu<\nu_\text{XUV}^{\min},\\
\beta_\text{XUV}\eta_\text{X}(r,\nu), & \nu_\text{XUV}^{\min}<\nu
   <\nu_\text{XUV}^{\max},\\
\eta_\text{X}(r,\nu), & \nu>\nu_\text{XUV}^{\max}.
\end{array}\right.
\end{equation}
For $\beta_\text{XUV}=0$ the additional XUV emission is not considered at all
and only additional X-ray ionization source is included,
whereas $\beta_\text{XUV}>1$ corresponds to enhanced source of the XUV emission.

\section{\ion{C}{v} ionization fraction and additional ionization sources}
\label{kadiskat}

The presence of lines of ions with higher degree of ionization (for example
\ion{C}{v}) is often used \citep[e.g., by][]{walcp} as an argument for the
existence of additional XUV or X-ray ionization source. Although it is generally
true that additional ionization source shifts the degree of ionization, due to
the complexity of the processes involved we can not claim that the presence of a
particular ion is caused exclusively by radiation at a chosen frequency.

As an example let us study the ionization ratio of \ion{C}{iv} and \ion{C}{v}.
Assuming that the population of the ground levels of these ions dominate, i.e.
that we can neglect population of excited states, the ionization balance between
the \ion{C}{iv} ion ($N_{4}$ is its number density) and \ion{C}{v} ion (number
density $N_{5}$) follows from the equations of statistical equilibrium
\citep{mihalas} as
\begin{equation}
\label{nlte45}
N_4 R_{45}-N_5R_{54}=0,
\end{equation}
where we took into account only the radiative ionization and recombination
(collisional transitions are neglected). The radiative ionization rate
\begin{subequations}
\label{zarion}
\begin{align}
R_{45} &= 4\pi \int_{\nu_4}^{\infty} \frac{\alpha_{4}(\nu)}{h\nu}
J(\nu)\,\de\nu,\\
\intertext{and the radiative recombination rate}
R_{54} &=4\pi \zav{\frac{N_4}{N_5}}^\ast
\int_{\nu_4}^{\infty} \frac{\alpha_4(\nu)}{h\nu}\hzav{\frac{2h\nu^3}{c^2}+
J(\nu)} e^{-\frac{h\nu}{kT}}
\,\de\nu,
\end{align}
\end{subequations}
where asterisk denotes LTE values and $\alpha_{4,\nu}$ is the photoionization
cross-section. Replacing the integrals in  Eq.~\eqref{zarion} with values at the
ionization frequency $\nu_4$ and taking into account that for a considered
spectral range ${2h\nu^3}/{c^2}\gg J(\nu)$, we derive from Eq.~\eqref{nlte45}
\begin{equation}
\frac{N_5}{N_4}= J(\nu_4) \frac{c^2}{2h\nu_4^3}\zav{\frac{N_5}{N_4}}^\ast
e^{\frac{h\nu_4}{kT}}.
\end{equation}
Using the Saha-Boltzmann equation, the fraction $({N_5}/{N_4})^*$ can be
eliminated and the latter equation can be further simplified (assuming unity
ionic partition function) to
\begin{equation}
\label{neelix}
\frac{N_5}{N_4}=\frac{c^2}{h^4\nu_4^3}\zav{2\pi m_\text{e}kT}^{3/2}
\frac{J(\nu_4)}{N_\text{e}},
\end{equation}
where $N_\text{e}$ is the electron density. From this equation it seems that
indeed the ionization ratio is directly proportional to the mean radiation
intensity $J$ at a given ionization frequency $\nu_4$. Using values appropriate
for the model 350-1 at roughly $2.3\,R_*$, we obtain in absence of additional
ionization sources for $\nu_4=1.6\times10^{16}\,\text{Hz}$,
$T=22\,000\,\text{K}$,
$N_\text{e}=3\times10^9\,\text{cm}^{-3}$, and $J(\nu_4)
=8\times10^{-15}\,\text{erg}\,\text{cm}^{-2}\,\text{s}^{-1}\,\text{Hz}$ from
Eq.~\eqref{neelix} the ionization ratio ${N_5}/{N_4}\approx7\times10^{-4}$. This
indicates low ionization fraction of \ion{C}{v} there. However, from our full
NLTE models, which consider reliable model ions, we obtain for the same location
${N_5}/{N_4}\approx0.2$.

The reason is that statistical equilibrium equations are quite complex in hot
star winds and their oversimplification using handy equations like
Eq.~\eqref{neelix} may lead to incorrect results. In the particular case of
\ion{C}{v} ionization fraction, the most important ionization process is not
that from the ground level, but from the less populated upper levels, which are
closely coupled with the ground level by a strong bound-bound transitions. This
information can be obtained \emph{only} solving the equations of statistical
equilibrium.

Consequently, a care has to be taken when making the conclusions about the
existence of additional ionization source just from the observations of
\ion{C}{v} lines. The same comment is valid also for \ion{N}{v} lines
\citep{pasam,nlteiii}.

\section{Influence of XUV radiation on \ion{P}{v} ionization fraction and wind
models}

\begin{figure}
\begin{center}
\resizebox{\hsize}{!}{\includegraphics{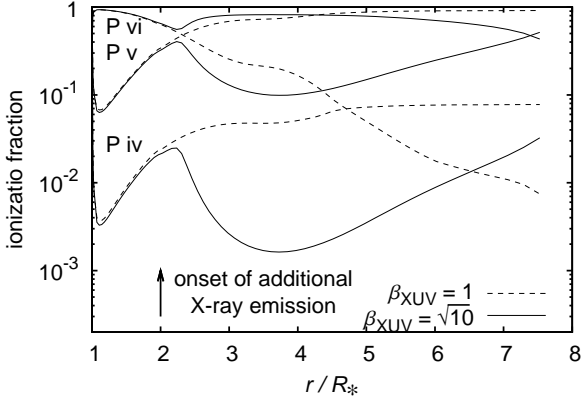}}
\end{center}
\caption[]{Ionization fraction of individual phosphorus ions in the model \mbox{400-1}
with different XUV sources}
\label{ion_P}
\end{figure}
\begin{figure}
\begin{center}
\resizebox{\hsize}{!}{\includegraphics{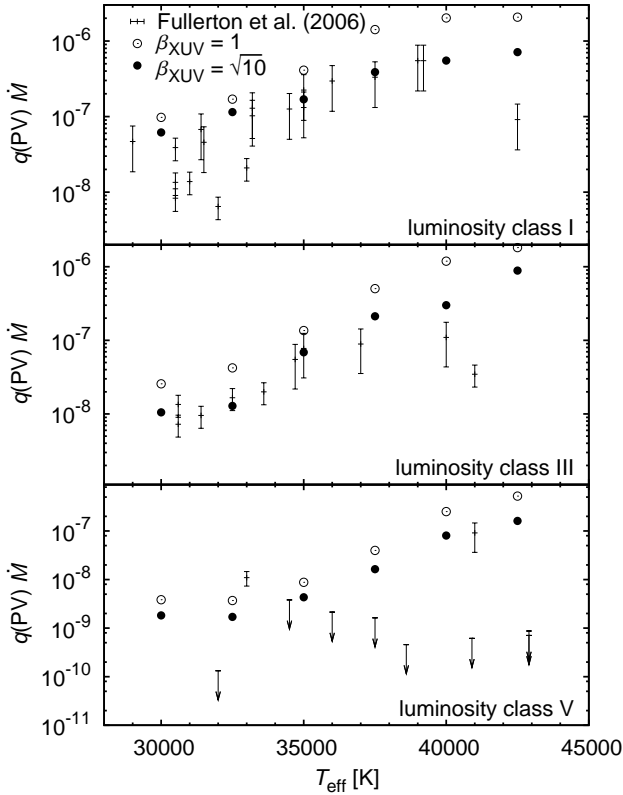}}
\end{center}
\caption[]{Comparison of predicted product of averaged \ion{P}{v} ionization
fraction and the wind mass-loss rate $q(\text{\ion{P}{v}})\dot M$ calculated
with $\beta_\text{XUV}=1$ (empty circles) and $\beta_\text{XUV}=\sqrt{10}$ (full
circles) with results derived from observations \citep[errorbars and upper
limits]{full} as a function of stellar effective temperature. The results for
different luminosity classes are separated into individual graphs.}
\label{ion_P_5}
\end{figure}

Let us now turn our attention to \ion{P}{v}. Without additional XUV sources, for
$\beta_\text{XUV}=0$ our models agree with the conclusions of \citet{nlteiii},
according to which the X-rays do not significantly influence the ionization
fraction of \ion{P}{v}. Only the ionization fraction of \ion{P}{vii} (and
\ion{P}{vi} in cooler stars) is significantly influenced by the X-rays, while
\ion{P}{v} stays to be one of the dominant ionization stages.

The inclusion of XUV radiation (with $\beta_\text{XUV}=1$, see Fig.~\ref{ion_P})
leads to a slight reduction of the \ion{P}{v} ionization fraction to roughly
$60\%$ of its value calculated without XUV radiation ($\beta_\text{XUV}=0$).
This indicates that a stronger XUV source may lead to a significant reduction of
\ion{P}{v} ionization fraction. Indeed, the \ion{P}{v} ionization fraction is
significantly reduced in the models with $\beta_\text{XUV}=\sqrt{10}$, as shown
in Fig.~\ref{ion_P_5}. Here we plot the product of \ion{P}{v} ionization
fraction and wind mass-loss rate $q(\text{\ion{P}{v}})\dot M$ averaged over
radii $2\leq r/R_*\leq\min(5,x_{\max})$ with $x_{\max}$ being the radius of the
outer model boundary (in units of $R_*$).

The comparison of predicted \ion{P}{v} ionization fractions and those derived
from observations by \citet{full} in Fig.~\ref{ion_P_5} supports the suggestion
of \citet{walcp} that enhanced source of XUV radiation leads to a reduction of
the \ion{P}{v} ionization fraction. There is a good agreement between the
product $q(\text{\ion{P}{v}})\dot M$ for the parameter
$\beta_\text{XUV}=\sqrt{10}$ and the observational one \citep[determined
by][]{full} for supergiants and giants. On the other hand, the results for
main-sequence stars are ambiguous. While for main-sequence stars with detected
\ion{P}{v} line there is a good agreement with observations even without
additional X-ray/XUV source, for stars with only upper limit of
$q(\text{\ion{P}{v}})\dot M$ available even a very strong X-ray/XUV source does
not bring the predictions and observations into agreement. Such disagreement can
be a signature of a "weak wind problem" \citep[e.g.][]{bourak,martclump}.

\citet{walcp} argued that due to a unique distribution of XUV radiation, where
the strongest XUV emission lines have energies lower than the \ion{S}{v} edge
(72.7~eV), the XUV radiation affects the ionization fraction of \ion{P}{v} and
not that of \ion{S}{v}. This effect might not be fully included in our models,
where the XUV emission lines are summed over a corresponding wavelength region.
This ensures that the total line emissivity is properly taken into account, but
some subtle effects of line distribution may be missing. To test the effect of
XUV line distribution, we calculated additional models with XUV emission present
only for energies lower than 72~eV. These models confirm that with a convenient
distribution of XUV emission the influence of XUV radiation on \ion{P}{v}
ionization fraction is more significant than on \ion{S}{v} (roughly by a factor
of 2). However, even the ionization fraction of \ion{S}{v} is affected by XUV
radiation with energies lower than the ionization energy of \ion{S}{v} due to
the ionization from higher levels of \ion{S}{v}, especially from the relatively
strongly populated second level 3s3p~$^3$P with a ionization energy 62.4~eV.
This is a similar situation to \ion{C}{v} ionization fraction discussed in
Sect.~\ref{kadiskat}.

\begin{figure}
\begin{center}
\resizebox{\hsize}{!}{\includegraphics{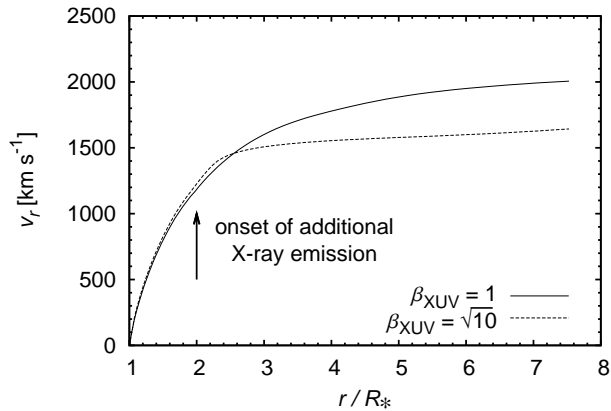}}
\end{center}
\caption[]{Influence of XUV radiation on the wind radial velocity in the model
400-1.}
\label{400_1vr}
\end{figure}

However, despite the promising results derived for \ion{P}{v} in giants and
supergiants, a detailed inspection of our models (which consistently include
\emph{all} possible driving ions) shows that there exists one additional (and
natural) effect of the XUV radiation. Phosphorus is a trace element and changes
of its ionization balance have only negligible effects on the radiation force.
However, also many ionization states of non-trace elements, which are the wind
drivers, are depopulated by XUV radiation. These ions include \ion{C}{iv}
(ionization energy $E_\text{ion}=64.5\,\text{eV}$), \ion{N}{iv}
($E_\text{ion}=77.5\,\text{eV}$), \ion{O}{iii} ($E_\text{ion}=54.9\,\text{eV}$),
\ion{O}{iv} ($E_\text{ion}=77.4\,\text{eV}$), and \ion{Si}{iv}
($E_\text{ion}=45.1\,\text{eV}$). This causes drastic changes in the line force
accelerating the whole wind. Consequently, the wind becomes overionized and the
radiation driving by higher ionization states becomes inefficient due to their
insufficient line opacity. This causes significant lowering of the radiative
force. Consequently, the wind overionization leads to wind stagnation that
starts roughly at the radius where the XUV radiation is switched on in our
models, i.e., at $r=2R_*$ (see Fig.~\ref{400_1vr}). In our calculations models
of giants and supergiants of all spectral types and main sequence stars with
$T_\text{eff}\gtrsim35\,000\,\text{K}$ are subject to this effect.

\begin{figure}
\begin{center}
\resizebox{\hsize}{!}{\includegraphics{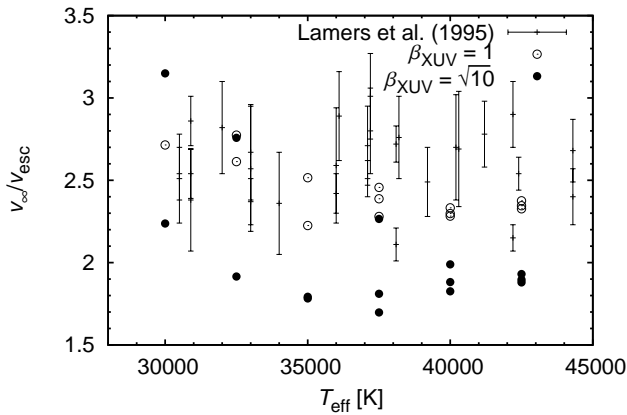}}
\end{center}
\caption[]{Comparison of predicted ratio of the wind terminal velocity and
escape velocity calculated with $\beta_\text{XUV}=1$ (empty circles) and
$\beta_\text{XUV}=\sqrt{10}$ (full circles) with results derived from
observed $v_\infty$ \citep{lsl}.}
\label{vnekvuni}
\end{figure}

Such stagnation would be observationally manifested by a decrease of the wind
terminal velocity. In Fig.~\ref{vnekvuni} we plot the ratio of the wind terminal
velocity to the escape velocity $v_\infty/v_\text{esc}$ from our models in
comparison with observational results. For a weak additional XUV source
$\beta_\text{XUV}=1$ the predicted $v_\infty/v_\text{esc}$ nicely reproduces the
observed results \citep{nlteiii}. However, for a stronger XUV source
$\beta_\text{XUV}=\sqrt{10}$ that reproduces the observed \ion{P}{v} ionization
fractions, the predicted $v_\infty/v_\text{esc}$ is significantly lower than
that based on observational results. This disagreement could be in principle
partially compensated in models with an onset of additional XUV emission farther
from the stellar surface, however this is supported neither by models
\citep{felpulpal} nor by the observations
\citep[e.g.,][]{celakoza,leucochce,coh}. By contrast, the onset of XUV emission
closer to the star suggested by observations leads to even lower wind terminal
velocity, which does not agree with observations.

\section{Discussion: wind mass-loss rates -- theory meets observations?}

The question of correct mass-loss rate determination is the most important one
of any wind theory and observation. Despite a significant progress in line
driven wind theory and especially in our understanding of the radiative transfer
in structured winds in recent years \citep{sund,clres1}, the answer to this
question is not unambiguously solved. Concerning the \ion{P}{v} ionization
fractions studied here, a reduction of mass-loss rate predictions by a factor of
about 3 is necessary to bring the observation of giants and supergiants and
theory into agreement (assuming $\beta_\text{XUV}=1$ not to violate the observed
terminal velocities).

\subsection{\ion{P}{v} ionization fractions with XUV emission and clumping}

\citet{pulamko} proposed that the problem of weak \ion{P}{v} lines is caused by
microclumping \citep[see also][]{prvnifosfor,sanya}. High density inside the
clumps favors the recombination leading to decrease of \ion{P}{v} ionization
fraction. We tested if the combination of microclumping and additional XUV
source does not bring observations and theory into agreement. For this purpose
we calculated other models, in which we both allowed for additional XUV emission
and took into account the influence of higher wind density in the clumps on the
NLTE equations \citep[as in][]{sanya}. Within this microclumping approach the
inhomogeneities directly affect the ionization equilibrium only and do not
influence the radiative transfer due to porosity. We assume that the clumping
starts above the critical point at the same radius as the additional X-ray
emission, i.e., at $2\,R_*$. In this case the clumping does not affect the
predicted mass-loss rates \citep[cf.][]{sanya}.

Our models show that clumping does not bring observations and theory into
agreement. Clumping has an opposite effect than additional XUV source decreasing
the wind ionization. Therefore, with microclumping (and XUV emission) the
phosphorus ionization fractions disagree with values derived from observations,
while the higher radiative force (due to lower ionization) provides better
agreement with observed terminal velocities. We were unable to find such
combination of parameters (describing additional XUV emission and microclumping)
that would provide both phosphorus ionization fractions and terminal velocities
in agreement with observations. On the other hand, the porosity, i.e., the
effect of wind structure on the line formation \citep{lidaarchiv,sund,taumy} is
the more promising one for the explanation of the remaining discrepancy between
theoretical and observed \ion{P}{v} line profiles.

\subsection{Mass-loss rates from X--ray
diagnostics}

\begin{table*}
\caption{Wind mass-loss rates estimated for individual stars from X-ray
diagnostics compared with our prediction after Eq.~\eqref{tricko}.}
\centering
\begin{tabular}{lcccccl}
\hline
\hline
Star & Sp. type & $\log({L}/{1\,\text{L}_\odot})$ & \multicolumn{2}{c}{$\dot M$
[$10^{-6}\,$M$_\odot\,$year$^{-1}$]} & Source (luminosities and X-ray
%
\\ &&&X-ray diagnostics &prediction& mass-loss rates) \\
\hline
HD 93129A &  O2 If & 6.17 & $6.8\pm2.5$ &$9.2$& \citet{coh}, \citet{rep}\\
$\zeta$ Pup & O4If & 5.86 & $3.5\pm0.3$ &$2.2$ & \citet{cohzet}, \citet{pulchuch}\\
HD 93250 & 
O4IIIf
& 5.95 & $1.4\pm0.5$ & $6.2$ & \citet{gagne}\\
9 Sgr & O4V & 5.67 & 0.34 & $1.8$ & \citet{cohact}, \citet{okali}\\
\hline
\end{tabular}
\label{tabulka}
\end{table*}

The situation is a bit different in the case of the mass-loss rate derived from
X-ray line profiles than in the case of \ion{P}{v} line profiles The shape of
X-ray line profiles may also imply low mass-loss rates in hot stars
\citep{celakoza}. Contrary to \ion{P}{v} line profiles, the microclumping does
not affect the shape of X-ray line profiles. However, the influence of porosity
is still debated \citep{taulida,tauowo}. Moreover, these determinations involve
some simplifying approximations, for example, the constant ionization structure
of the ambient cool wind.

In Table~\ref{tabulka} we compare the mass-loss rates calculated using formula
Eq.~\eqref{tricko} with those derived from X-ray line profiles
and from continuum X-ray absorption in the case of HD 
93250.
The comparison
may be biased, because 
HD 
93250
and 9 Sgr are binaries and show peculiar features, including
enhanced or non-thermal radio emission or coliding winds
\citep{lechko,rablowal,infersane,gagne,cekali}.
Note also that the mass-loss rate in
the case of $\zeta$ Pup was derived from observations assuming non-solar
chemical composition \citep{cohzet}, while our models assume solar chemical
composition. 

The results shown in Table~\ref{tabulka} show that there is a good agreement
between wind mass-loss rates derived from observations and theory for the two
supergiants, while for the other two
non-supergiant stars
predicted rates overestimate the observed ones by a factor of about 5.
This result may reflect a
similar
dichotomy between supergiant and main-sequence mass-loss rates 
%
found from
\ion{P}{v} ionization fractions (see Fig.~\ref{ion_P_5}).

\subsection{$\rho^2$ diagnostics as upper limit to the mass-loss rates}

The H$\alpha$ emission line and infrared (or radio) continua belong to
traditional mass-loss rate indicators. However, the amount of H$\alpha$ emission
or infrared excess is not directly proportional to the wind density $\rho$, but
to its square $\rho^2$ \citep[e.g.,][]{pulamko}. The structured (clumped) wind
with lower mass-loss rate may mimic spectral features of the homogeneous wind
with higher mass-loss rate. Consequently, the H$\alpha$ emission line or
infrared continua may provide only upper limits to the mass-loss rate
\citep{pulchuch}, since they give values which correspond to a homogeneous wind.
The predicted mass-loss rates should be lower than that derived from $\rho^2$
diagnostics (i.e., H$\alpha$ emission line or infrared continua).

\begin{figure}
\begin{center}
\resizebox{\hsize}{!}{\includegraphics{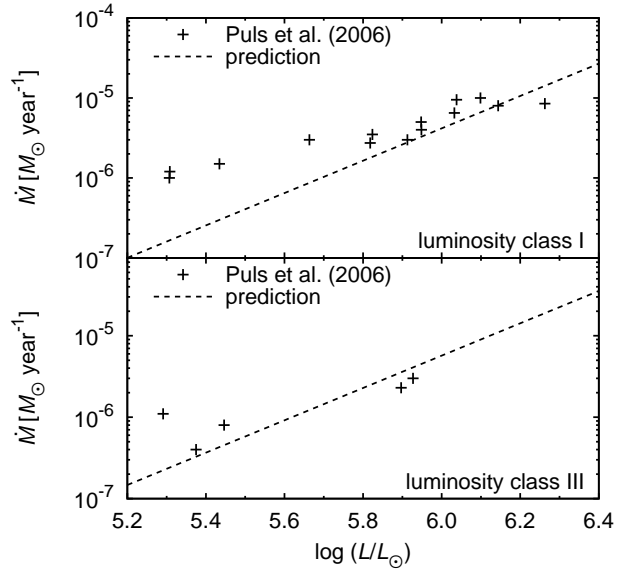}}
\end{center}
\caption[]{Comparison of upper limits to the mass-loss rates from $\rho^2$
diagnostics \citep{pulchuch} with our
predicted dependence of mass-loss rates
Eq.~\eqref{tricko} as a function of stellar luminosity.}
\label{chuchval}
\end{figure}

In Fig.~\ref{chuchval} we plot the predicted mass-loss rates calculated using
Eq.~\eqref{tricko} as a function of stellar luminosity in comparison with
observational upper limits derived from $\rho^2$ diagnostics by
\citet{pulchuch}. The luminosity of stars from this observational sample was
derived using the stellar parameters taken from \citet{pulchuch}. For
supergiants with higher luminosity $\log(L/L_\odot)\gtrsim5.8$ and for giants
the mass-loss rates derived from observations are on average only slightly
higher than predicted ones. This indicates only medium level of clumping in
these stars. On the other hand, the mass-loss rates derived from observations
are significantly higher than predictions for low-luminosity supergiants
$\log(L/L_\odot)\lesssim5.8$, pointing to a more significant level of clumping
in their winds. However, in general the predicted mass-loss rates are not higher
than those derived from the $\rho^2$ diagnostics. Consequently, even in this
case the predictions do not contradict the observations.

\section{Conclusions}

We tested the influence of XUV radiation on \ion{P}{v} ionization fractions in
hot star winds. Using hydrodynamical NLTE wind modelling we confirmed the
conclusion of \citet{walcp} that the XUV radiation may decrease the \ion{P}{v}
ionization fraction. As the result of various ionization and recombination
processes, the \ion{P}{v} ionization fraction in hot star winds does not come
close to the value of 1, implying that the role of the \ion{P}{v} lines as
mass-loss rate indicators in hot star winds is not so straightforward. On the
other hand, the large amount of XUV radiation necessary to significantly lower
the \ion{P}{v} ionization fraction to the values implied by observations leads
to a decrease of the wind terminal velocity due to inefficient line driving.
This contradicts the observations. Moreover, the wind clumping has an opposite
effect than the XUV emission decreasing the wind ionization. Consequently, it is
unlikely that the effect of lowering the \ion{P}{v} ionization fraction by XUV
radiation alone can bring the theory and observations in accord.

We also provide a useful mass-loss rate formula and compare it with other
mass-loss rate diagnostics. We show that for supergiants and giants this formula
passes two important observational tests against mass-loss rates derived from
X-ray line profiles and $\rho^2$ diagnostics. This supports the reliability of
mass-loss rates predictions derived from modern wind codes for luminous hot
stars.

For main-sequence stars the predicted mass-loss rates are significantly higher
than those inferred from the \ion{P}{v} or X-ray lines. This may indicate that
the "weak wind problem" recently detected in low-luminosity main-sequence stars
occurs to some extent
also for the stars with higher luminosity. While the explanation of this
"weak wind problem" for low-luminosity main-sequence stars is at hand (too large
cooling length, \citealt{luciebila}, \citealt{martclump}, \citealt{cobecru},
\citealt{nlteiii}, \citealt{lucyjakomy}, or influence of X-rays on the mass-loss
rate, \citealt{nemajpravdu}) such explanation valid for all main-sequence stars
with any luminosity is currently missing.

\section*{Acknowledgements}
We thank Prof.~Achim Feldmeier for providing us results of his simulations
without which this research would not be possible, and to Dr.~J.~Puls for
providing us atomic data for phosphorus. This work was supported by grant GA
\v{C}R 205/08/0003. The access to the METACentrum (super)computing facilities
provided under the research intent MSM6383917201 is also acknowledged. The
Astronomical Institute Ond\v{r}ejov is supported by the project RVO:67985815.

\newcommand{\jezero}[1]{in C.~Robert, N.~St-Louis, \& L.~Drissen eds.,
	ASP Conf. Ser., Four Decades of Research on Massive Stars,
	Astron. Soc. Pacific, San Francisco, #1}

\end{document}